# Author Name Disambiguation in Bibliographic Databases: A Survey

**Muhammad Shoaib[a], Ali Daud[b]*, Tehmina Amjad[c]**

[a]Department of Computer Science, Comsats University, Sahiwal Campus, Pakistan

[b]Department of Information Systems and Technology, College of Computer Science and Engineering, University of Jeddah, Saudi Arabia

[c]Department of Computer Science and Software Engineering, International Islamic University, Islamabad, Pakistan

shoaibIshaaqiiu@yahoo.com, ali_msdb@hotmail.com, tehminaamjad@iiu.edu.pk

**Abstract:** Entity resolution is a challenging and hot research area in the field of Information Systems since last decade. Author Name Disambiguation (AND) in Bibliographic Databases (BD) like DBLP[1], Citeseer[2], and Scopus[3] is a specialized field of entity resolution. Given many citations of underlying authors, the AND task is to find which citations belong to the same author. In this survey, we start with three basic AND problems, followed by need for solution and challenges. A generic, five-step framework is provided for handling AND issues. These steps are; (1) Preparation of dataset (2) Selection of publication attributes (3) Selection of similarity metrics (4) Selection of models and (5) Clustering Performance evaluation. Categorization and elaboration of similarity metrics and methods are also provided. Finally, future directions and recommendations are given for this dynamic area of research.

**Keywords:** Entity Resolution, Author Name Disambiguation (AND), Bibliographic Databases (BD), Similarity Metrics, Methods

## 1. Introduction

Entity resolution has attracted the attention of information system researchers for a long time now. AND in BD is hot issue and is a specialized filed of entity resolution. Author name disambiguation is the process of distinguishing authors with similar names from each other. The bibliographic databases include a large amount of data from co-author networks and digital libraries. Authors or researchers can have similar names, can have multiple ways of writing their full names, or different authors can share multiple names. These situations arise the ambiguity for the methods that need the publications metadata for ranking, or evaluating the authors [1] [2]. The disambiguation methods are not only required in co-author networks, but are also significant in fields like spam filtering [3]-[5]. The search engines like Google[4] facilitates the users in searching web pages automatically. The name queries are approximately 5-10% of all queries [6]. Further, it is estimated that the 300 most common male names are used by more than 114 million people in the United States [7]. Search engines usually treat the name queries as normal keyword search, and do not pay any special attention towards their possible ambiguity. For example, when searching for Tehmina Amjad on Google, it shows 129,000 web pages containing similar names. Out of these pages only a small portion is relevant to the intended Tehmina Amjad. This is because of the data on Internet is of heterogeneous nature.

In BD, it is necessary to uniquely identify the work of one researcher from another, and this process is known as AND. Formally, a bibliographic database is an organized digital store of citations to research publications, patents, books and news articles, etc. It stores the metadata of the publications. Examples of commonly used BD are: DBLP [8], CiteSeer [9], MEDLINE[5] and Google Scholar[6]. An AND method that best fits to a bibliographic dataset may not be suitable for other datasets. The reason behind is that they differ in their metadata schema. Most of the methods fall in either supervised learning or un-supervised learning or combination of the two.

[1] http://www.informatik.uni-trier.de/~ley/db/
[2] http://citeseer.ist.psu.edu/
[3] http://www.scopus.com/home.url
[4]http:// www.google.com
[5] www.ncbi.nlm.nih.gov
[6] scholar.google.com

Smalheiser and Torvik [10] have provided detailed literature survey of methods for AND but their work has many short comings, such as, a general framework is not provided, similarity metrics and methods are not explained category-wise in detail.

Our contributions in this work are as follows

(1) Proposal of a general framework for AND

(2) Categorization and elaboration of similarity metrics which are the main focus of researchers in AND to find the resemblance among citations and

(3) Categorization of methods used to handle AND task into five types with elaboration of works falling under each category in chronological order.

Rest of the paper is organized as follows. Section 2 describes AND task and related concepts. Section 3 provides a general framework based on most of the methods used in the past. Section 4 is about the commonly used datasets to perform AND. Section 5 is about the similarity estimation metrics. Section 6 categorizes the methods employed for AND, and explains categories in chronological order. Section 7 explains how to compare different methods and some future directions and recommendations are suggested in section 8. Finally, section 9 concludes this paper.

## 2. Author Name Disambiguation in Bibliographic Databases ($AND_{BD}$)

Resolving the name ambiguity in Bibliographic Databases is called $AND_{BD}$. In literature many terms are used for this problem like name disambiguation [11][12], object distinction [13], mixed and split citation [14], author disambiguation [15] and entity resolution [16][17]. $AND_{BD}$ problems can be divided into three categories. Before discussing $AND_{BD}$ problem categories through intuitive examples, some related basic concepts are provided.

Publication: A publication means the research work/article/paper of an author or group of authors working together published at any venue (journal, conference or workshop).

Citations: The number of times a publication is cited/referenced by other publications.

References: It is the list of references given at the end of a publication.

Ambiguous Author name(s): A name that is either shared by multiple authors or multiple variant names of a single author. Let A be the ambiguous author name shared by k number of unique authors, say, $a_1$, $a_2$,… ,$a_k$. Further let $a_i$ is an author represented by m number of various names, say, $n_1$, $n_2$,…, $n_m$. In this article, we use “ambiguous author name”, “ambiguous author” and “ambiguous name”, interchangeably.

### 2.1. Problem Categories

#### 2.1.1. Synonymy/Name Variant Problem

The problem of Synonymy arises when an author has variations or abbreviations in his/her name in the citations. For example, the author name “Malik Sikandar Hayat Khiyal” is also written as “Sikandar Hayat” in citations of the publications. The DBLP treats them as two different authors and divides his publications between two names. In literature, this problem is also referred to as name variant problem [17][18], entity resolution problem [16], split citation problem [14]**Error! Reference source not found.** and aliasing problem [19].

#### 2.1.2. Polysemy/Name Sharing Problem

The problem of Polysemy arises when multiple authors share the same name label in multiple citations. For example, “Guilin Chen” and “Guangyu Chen” write their names as “G. Chen” in their publications. It is quite possible that a full name of an author is shared by multiple authors. Bibliographic databases may treat these different authors as a single author. Resultantly, on querying the database for such ambiguous names, it may list all publications under a single person’s name. On querying DBLP against author name “Michael Johnson” it lists 32 publications these are actually from five different people [17]. In literature there are various names of this problem such as name disambiguation [11][12][20], object distinction [13], mixed citation [14], author disambiguation [15] and the common name problem [17].

#### 2.1.3. Name Mixing Problem

Shu et al. [17] introduced another type of name disambiguation problem and referred to it as name mixing problem. If multiple persons share multiple names it is called the name mixing problem. The two problems discussed above may occur simultaneously and cause the name mixing problem.

Typographical mistakes also cause the name ambiguity. Treeratpituk and Giles [19] consider the typographical mistakes in names as a separate name disambiguation problem. These problems may arise due to use of abbreviations, spelling mistakes; and occasionally using caste or family name at the end or in the beginning of names. L. Branting [21] has discussed nine different types of name variations.

### 2.2. Need for the Solution

Name ambiguity may cause incorrect authorship identification in literary works resulting in improper credit attribution to the authors. AND is a basic and compulsory step for performing bibliometric and scientometric analyses. Disambiguating authors may help establish precise; author profiles, co-author networks, and citation networks. In academic digital libraries, disambiguating author names is necessary for the following reasons.

- Users are interested in finding papers written by a particular researcher [98]
- Research communities and institutions can track the achievements of their researchers [99]
- It also helps in expert finding from which publishers can easily find paper reviewers [100]

### 2.3. Challenges involved in AND

There are certain challenges that are involved in AND, some of which are highlighted in the following.

- Lack of identifying information: The identifier metadata are either incomplete or not available at all.
- Multi-directional problem: Multi-disciplinary papers authored by multiple researchers from multiple institutions (nationwide or world-wide) may cause 'multiple entities disambiguation' problem.
- Less number of papers by most of the authors: The machine learning techniques used for AND give better results when reasonable number of examples is available. This is only possible when the individual authors have produced many papers. In MEDLINE almost 46% of the authors have written only one paper **Error! Reference source not found.**. The authors having one or few papers are big hindrance for proposing precise machine learning techniques.
- Heterogeneous nature of BD: The BD are heterogeneous in many ways, like: schema heterogeneity, discipline heterogeneity, language heterogeneity and attributes heterogeneity etc.
- Non-serious attitude of the authors: Sometimes the authors are reluctant in registering a universal identification system like UAI_Sys [22] or [23] or making consolidated profiles.
- Economic issue: The construction of such a database that can accommodate and manage the world-wide researchers' community including all the disciplines, nations and languages is not only economically unfeasible but also probably impossible.
- Ownership issue: While testing the algorithm for AND sometimes the confirmation of original author becomes doubtful.

### 2.4. Is Unique Identifier for an Author a Viable Solution?

One may think that unique identifiers, say, Author Identification Number (AID), can be simple and reliable solution for this problem. Dervos et al. [22] proposed UAI_Sys in which an author can register himself/herself by entering his/her metadata information. The UAI_Sys in return assigns a 16 digit unique code to the author. ORCID [23] is also a similar attempt for the same purpose, it issues a 16 characters alphanumeric code to the researcher to uniquely identify them. It offers a permanent identity for people, just like the ones issued for content-related entities on digital networks by digital object identifiers Although it seems possible apparently, however, there are so many issues discussed in this section that are very difficult to address and implement.

In Dervos et al. [22] project it is expected that authors would remember their passwords and UAIs. Researchers do not pay attention to remember such lengthy codes. Further, all the co-authors are also bounded to be registered with the universal bibliographic database. A large number of authors may produce 2 or 3 papers in their whole life. Such casual researchers take least interest to be registered in the database. It is not only the casual researchers but regular researchers (who produce reasonable number of research papers) may also provide wrong metadata information to the system. Sometimes it is too difficult to convince a researcher to be habitual to welcome new technologies. They may resist giving up the previous practices and adopting the new ones.

If such a database is developed, ideally it should accommodate all the research areas, languages, states and all types of publications. Such a database seems not to be economical as it demands not only one-time expenses (developing cost) but also huge running expenses including staff salaries, maintenance and security of the database, and handling the user queries.

### 2.5. Mathematical Notations

Table 1 provides the mathematical notations used in this paper.

Table 1: Mathematical notations

| Symbols | Sets | Description |
|---|---|---|
| A | A= {$a_1$, $a_2$, …, $a_k$}, where $a_i$ is the *ith* author; *k* is no. of unique authors sharing an ambiguous name | Set of authors/persons sharing an ambiguous name |
| D | D= {$d_1$, $d_2$, …, $d_d$} | Set of documents in a dataset |
| P | P= {$P_1$, $P_2$, …, $P_p$} | Set of publications/documents associated to an ambiguous author/name |
| K | | No. of clusters = No. of unique authors associated to an ambiguous name |
| V | V = {$v_1$, $v_2$, …, $v_v$}, where v is the number of vertices | Set of vertices in a graph |
| E | E = {$e_1$, $e_2$, …, $e_e$}, where e is the number of vertices | Set of edges in a graph |
| N | | Number of unique authors |
| *w* | | Set of words |
| *t* | | Term, can be a word or set of words |

# 3. AND$_{BD}$ Process

In this section, we describe the general process of AND$_{BD}$. We do not follow the process exploited by any particular research work. We actually provide the common steps involved in AND$_{BD}$ process. The purpose of this section is to help readers comprehend AND$_{BD}$ task more easily and clearly. Figure 1 is the block diagram of the AND$_{BD}$ process.

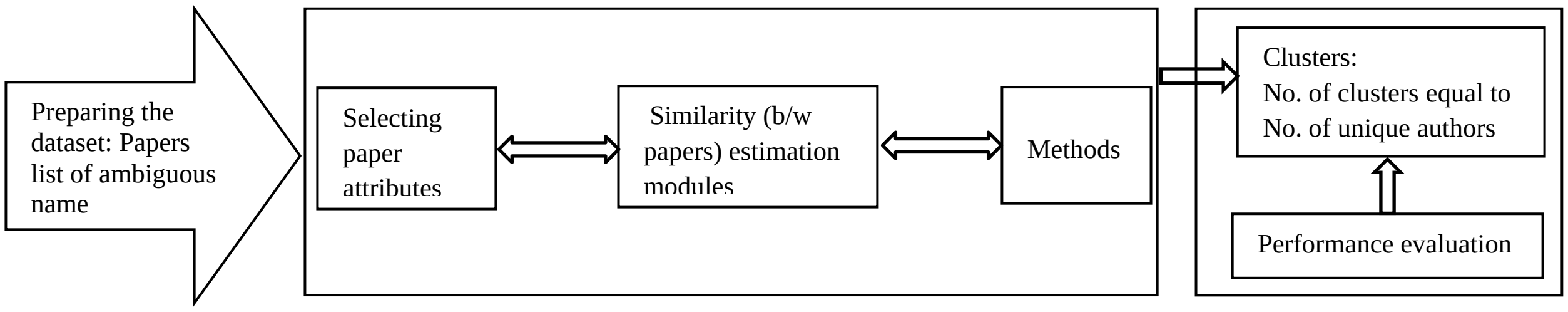

Figure 1: AND$_{BD}$ process

## 3.1. Preparing the Dataset

For AND a BD is used. The whole database is normally too large to analyze, within a limited time. To avoid killing time in query processing in real life databases, a tiny dataset is either selected from a functional BD or prepared from scratch normally by crawling the web pages of ambiguous authors. For example, Han et al. [24] exploit two datasets, one for 15 different "J. Anderson"s, and the other for 11 unique "J. Smith"s; while Wang et al. [25] used a dataset containing 16 ambiguous names comprising 241 unique authors. Preprocessing in name disambiguation usually includes blocking, stop-word removal, and stemming [26]. Stop-word removal and stemming steps are required for the title words of publications and venues. Blocking step is performed to group together the authors with ambiguous names. Disambiguation operations are performed within each ambiguous group to avoid useless comparisons and operations involving non-ambiguous authors.

## 3.2. Selecting the Publication Attributes

It is always desirable to utilize as many attributes of the publications as available though only useful ones are considered. All BD do not provide same number and type of attributes. But three common attributes: co-authors, publication title and venue; are available in almost all of them. We name these three attributes as triplet attributes. Most of the studies like [24] uses only triplet attributes, [17] exploits triplet attributes plus topic similarity. Some methods like [25][27] take advantage of indirect co-authors, feedback, co-web and publication year along with triplet attributes. Torvik et al. [28] propose eight different attributes: (1) middle initial, (2) suffix (e.g., Prof. or II), (3) full name, (4) language, (5) number of common co-authors, (6) number of common title words, (7) number of common affiliation words and (8) number of common Medical Subject Headings (MeSH) words. As we add more and more attributes, usually the accuracy increases a bit at the cost of time complexity. In AND time complexity is not much cumbersome, however, unavailability of reasonable number of distinguishing attributes is a bottleneck.

### 3.3. Selecting the Similarity Estimators

After the selection of available attributes, the most technical task is to select a proper similarity estimator for the attributes. Almost all the methods in AND, work on the notion that, more the similarity values among the attributes of the two citations, more it is plausible that they belong to the same author. The main focus of the proposed similarity estimators is always to estimate the optimum similarity value among the attributes of the two papers. Various similarity estimators for each type of attribute are exploited by the researchers. For example, Shu et al. [17] used edit distance of two strings for co-author attribute, cosine similarity measure for title and venue attributes, and Latent Dirichlet Allocation (LDA) [29] topic model for semantic topic similarity.

### 3.4. Selecting the Models

In this study, we categorized the AND methods into five types (1) supervised learning (2) unsupervised learning (3) semi-supervised learning (4) graph based and (5) ontology based. Supervised learning models perform classification, unsupervised learning methods perform clustering and semi-supervised models are combination of both supervised and unsupervised methods. Graph based methods exploit links and ontology based methods exploit semantics based relationships between entities. The purpose of all methods is to separate the publications of a unique author into a unique class/cluster. A large number of methods are available, so first of all one must decide which type of method will be employed. Pros and cons of each alternate are kept in mind before applying the method. One can think to devise his/her own new method as well. SVM and decision tree algorithm C4.5 classifiers are widely used classification models in AND. On the other hand, random forests, spectral clustering and DBSCAN are popular clustering models.

### 3.5. Measuring the Performance

The performance of the method used is measured using different performance metrics. Precision, recall and F-measure are very common performance metrics used for evaluation of AND methods.

## 4. Datasets

The well-known BD like DBLP, MEDLINE, DBComp, Scopus and CiteSeer have been widely utilized by the researchers for AND. DBLP is the most widely used database for this purpose. Its basic reason, perhaps, is that the publication records in DBLP are represented in a well-structured format, i.e., XML. The basic issue faced by the researchers is how to measure the performance of the proposed method with standard/huge databases. For this purpose, they pick few ambiguous names from the database along with their publications and other discriminative attributes, and investigate the performance of their proposed method.

For example, Han et al. [24] exploited two types of datasets: (1) Collected manually from web by querying Google, and (2) selected ambiguous names from DBLP. The first dataset consists of two ambiguous names “J. Anderson” and “J. Smith”. “J. Anderson”. Part of the dataset consists of 15 unique authors who share the same name, and 229 publications; “J. Smith” is shared by 11 different authors whose total publications are 338. “J. Anderson” part of the first dataset is shown in Table 2. Tables, 2, 3 and 4 show some examples of name ambiguity. We can see from Table 2 that there are 15 different people whose first name are James and Last names are Anderson. However, they have different middle initial. All these names can appear in a publication as J. Anderson, and it needs to be resolved that which J. Anderson is actually intended. The second dataset consists of 9 ambiguous names with each having more than 10 name variations, as shown in Table 3. These datasets, later on, were used by many other works like [11][30] etc.

Ferreira et al. [31] also used two datasets. They collected records from DBLP and DBComp. The statistics are given in Table 4. Many other studies like [11][30][32][33] have used these dataset with some variations. Reuther [34] investigated the existing test collections and proposed three new test collections to resolve the name variant problem.

Table 2: “J. Anderson” part of first dataset used by Han et al. [24]

| Full Name | Affiliation | No. of Pubs | Full Name | Affiliation | No. of Pubs |
|---|---|---|---|---|---|
| James Nicholas Anderson | UK Edinburgh | 8 | James D. Anderson | Univ. of Toronto | 5 |
| James E. Anderson | Boston College | 14 | James P. Anderson | N/A | 3 |
| James A. Anderson | Brown University | 3 | James M. Anderson | N/A | 5 |
| James B. Anderson | Penn. State Univ | 6 | James Anderson | UK | 19 |
| James B. Anderson | Univ. of Toronto | 21 | James W. Anderson | Univ. of KY | 10 |
| James B. Anderson | Univ. of Florida | 17 | Jim Anderson | Univ. of Southampton | 20 |
| James H. Anderson | Univ. of North Carolina | 54 | Jim V. Anderson | Virginia Tech Univ. | 40 |
| James H. Anderson | Stanford Univ. | 4 | | | |

Table 3: Second dataset used by Han et al. [24]

| **Ambiguous Names** | **Name Variations** | **No. of Pubs** | **Ambiguous Names** | **Name Variations** | **No. of Pubs** |
|---|---|---|---|---|---|
| S Lee | 35 | 467 | C Lee | 18 | 152 |
| J Lee | 33 | 330 | A Gupta | 16 | 332 |
| J Kim | 25 | 239 | J Chen | 13 | 174 |
| Y Chen | 24 | 201 | H Kim | 11 | 120 |
| S Kim | 20 | 181 | | | |

Table 4: Datasets used by Ferreira et al. [31]

| **DBLP** | | | **DBComp** | | |
|---|---|---|---|---|---|
| **Ambiguous Names** | **No. of Authors** | **No. of Pubs** | **Ambiguous Names** | **No. of Authors** | **No. of Pubs** |
| A. Gupta | 26 | 576 | A. Oliveira | 16 | 52 |
| A. Kumar | 14 | 243 | A. Silva | 32 | 64 |
| C. Chen | 60 | 798 | F. Silva | 20 | 26 |
| D. Johnson | 15 | 368 | J. Oliveira | 18 | 48 |
| J. Martin | 16 | 112 | J. Silva | 17 | 36 |
| J. Robinson | 12 | 171 | J. Souza | 11 | 35 |
| J. Smith | 29 | 921 | L. Silva | 18 | 33 |
| K. Tanaka | 10 | 280 | Silva | 16 | 21 |
| M. Brown | 13 | 153 | R. Santos | 16 | 20 |
| M. Jones | 13 | 260 | R. Silva | 20 | 28 |
| M. Miller | 12 | 405 | | | |

## 5. Similarity Metrics

Selecting an appropriate similarity metric/distance function is a technical and challenging task [35] in AND. It is advisable to employ the best fit similarity measure for each attribute of the publications. No single metric is the best fit for all the attributes. Cohen et al. [36] compared different similarity metrics for name matching and concluded that combination of metrics provide better results than any single metric. Most of the similarity measures do not make use of semantics of the publications and use syntactic characteristics only, so we categorize these metrics into two types (1) syntactic and (2) semantic similarity metrics.

### 5.1. Syntactic Similarity Metrics

The similarity metrics that match the strings exactly and do not care about synonymy and polysemy are syntactic similarity metrics. The similarity of two publications can be obtained by cosine, Euclidean, Manhattan, Jaccord, Jaro, Winker, TFIDF, etc. These metrics often outperform Levenshtein-distance-based techniques [36]. Besides these metrics many other measures like typewriter distance, Jaro-Winkler, Monge-Elkan, or phonetic distances can also be employed. The most used metrics of sub categories are (1) edit distance and (2) token based distance metrics of syntactic similarity.

*5.1.1. Edit Distance Metrics*

Distance functions map two strings $S_1$ and $S_2$ to a real number $r$, where a larger value of $r$ indicates greater distance or smaller similarity between $S_1$ and $S_2$. String distances are most useful for matching problems with little prior knowledge and/or ill-structured data [36]. Variety of edit distance functions are used in text mining tasks. The edit distance of two strings (names) is the minimum number of operations required to transform one string to the other. These operations include insertion, deletion and replacement of a character. A good comparison of name matching techniques is given in [36].

The most simple is Levenshtein distance [36] that assigns a unit cost to all edit operations. Monger-Elkan distance function [37] is more complex and well-tuned with particular cost parameters and is scaled to the interval (0, 1). It is a variant of the Smith-Waterman distance function [38] and assigns a relatively lower cost to a sequence of insertions or deletions.

Shu et al. [17], Bhattacharya and Getoor [16], Torvik et al. [28] and Smalheiser and Torvik [10] utilized edit distance like measures for measuring name distance of the co-authors of two citations. Shu et al. [17] applied rule based methodology along with edit distance.

A little bit similar metric, but not based on edit distance model is Jaro metric [39], which is based on the number and sequence of the common characters between the two strings [14][19][26]. A variant of this function is Jaro-Winkler [40], which exploits the length of the longest common prefix between $S_1$ and $S_2$ [14][19][26][41].

*5.1.2. Token based Distance Metrics*

Token based distance metrics compare words of the two strings S1 and S2 rather than the characters. Euclidean distance is commonly used for text clustering problems and similarity estimation [6][13][27][30][42]. Let $d_1$ and $d_2$ represent vectors of two documents then the Euclidean distance between the two documents can be calculated as:

$$\boldsymbol{DIST_E}(\mathbf{d_1}, \mathbf{d_2}) = \sqrt{\sum_{t=1}^{n} |\boldsymbol{w_{t1}} - \boldsymbol{w_{t2}}|^2} \dots \dots \dots \dots \dots \dots (\mathbf{1})$$

where, term frequency $t_i \in$ T and T = $\{t_1, . . . , t_n\}$.

Term Frequency Inverse Document Frequency (TFIDF) is the frequency of word *w* in an attribute of a publication, and IDF is the inverse of the fraction of words in the dataset that contains *w* and is used by [11][14][19][26][43][44]**Error! Reference source not found.**. Cohen et al. [36] considered a soft version of TFIDF in which similar tokens are also considered along with tokens in $S_1 \cap S_2$. Most of the research works like [14][15][17][24][25][26][27][31] use the cosine similarity that exploits TFIDF and vector space model (VSM) [45]. Normally this function is used for title and venue attributes. Although, it can be used for any attribute represented in the form of vectors. The documents are represented in vector space. Let $d_1$ and $d_2$ represent vectors of two documents then the cosine similarity between the two documents can be calculated as:

$$SIM_C(\mathrm{d_1}, \mathrm{d_2}) = Cosine\ \theta = \frac{\mathrm{d_1.d_2}}{|\mathrm{d_1}|.|\mathrm{d_2}|} \dots \dots \dots \dots \dots \dots (2)$$

Jaccard coefficient, also called the Tanimoto coefficient, is the ratio between the intersection and the union of the objects. It compares the sum weight of common terms to the sum weight of terms that are present in either of the two documents except the common terms [13][14][19][26][44]. Let $d_1$ and $d_2$ represent vectors of two documents. The Jaccard coefficient between the two documents is:

$$SIM_J(\mathrm{d_1}, \mathrm{d_2}) = \frac{\mathrm{d_1.d_2}}{|\mathrm{d_1}|^2 + |\mathrm{d_2}|^2 - \mathrm{d_1.d_2}} \dots \dots \dots \dots \dots \dots (3)$$

A document can also be considered as a probability distribution of terms in probability theory. The similarity between the two documents can be calculated by measuring the distance between the two corresponding probability distributions. Let $d_1$ and $d_2$ represent vectors of two documents, the KL divergence between the two distributions of words is calculated as:

$$D_{KL}(\mathrm{d_1} || \mathrm{d_2}) = \sum_{t=1}^{n} w_{t1}\ \mathrm{X}\ log\frac{w_{t1}}{w_{t2}} \dots \dots \dots \dots \dots \dots (4)$$

The KL divergence is not symmetric on the other hand average KL divergence is symmetric, that is why average KL divergence is more popular. The average weighted KL divergence from $d_i$ to $d_j$ is the same as that of from $d_j$ to $d_i$. This average weighting between two vectors of the two corresponding documents guarantees symmetry. For text documents, the average KL divergence between the two distributions of words is calculated as:

$$D_{AvgKL}(\mathrm{d_1} || \mathrm{d_2}) = \sum_{t=1}^{n} (\barwedge_1\ \mathrm{X}\ D\ (w_{t1}||w_t) + (\barwedge_2\ \mathrm{X}\ D\ (w_{t2}||w_t)) \dots \dots \dots \dots \dots \dots (5)$$

where, $\barwedge_1 = \frac{w_{t1}}{w_{t1} + w_{t2}}$, $\barwedge_2 = \frac{w_{t2}}{w_{t1} + w_{t2}}$ and $w_t = \barwedge_1\ X\ w_{t1} + \barwedge_2\ X\ w_{t2}$

## 5.2. Semantic Similarity Metrics

The above discussed measures help in estimating pair-wise similarities between the corresponding attributes of the publications. They usually exploit syntactic characteristics and are unable to utilize the Synonymy and Polysemy based semantics of publications. The topic models such as PLSA [46] and LDA [29] provide excellent ways to exploit semantics. A publication mostly contains multiple topics and it is important to find the topic similarity between the two publications. Generally, a topic is a semantically related probabilistic cluster of terms (words). Here, we describe LDA which can capture semantics in an unsupervised way. It is a generative probabilistic model for text corpora [29][47][48] at words and documents level. It assumes every document as a mixture of topics and every topic as a Dirichlet distribution over words in the vocabulary. It has been used for finding topic similarity among the publications [6][16][24]. Shu et al. [17] and Song et al. [6]**Error! Reference source not found.** extend the LDA model and applied it to AND. The probability of generating word *w* from document d is given as:

$$P(w|d,\theta,\Phi) = \sum_{z=1}^{T} P(w|z,\Phi_z)P(z|d,\theta_d) \ldots\ldots\ldots\ldots\ldots (6)$$

Where, $w$ is vector form of $d$, $z$ is topic and $\boldsymbol{\Theta_d}, \boldsymbol{\Phi_z}$ are multiple distributions over topics and over words specific to $z$, simultaneously.

## 6. Approaches for $AND_{BD}$

Much research work has been done on entity resolution in variety of research areas. In the field of databases, studies are made on merge/purge [49], record linkage [50]**Error! Reference source not found.**, duplicate record detection [51], data association [52] and database hardening [53] etc. In Natural Language Processing (NLP), Cross-Document Co-Reference [54] methodologies and name matching algorithms [21] are designed. In BD, several methods or models are employed, such as, citation matching [55], k-way spectral clustering [11], social network similarity [12], mixed and split citation [14], Latent Topic Model [17], latent Dirichlet model [16], Random Forests [19], Graph-based GHOST [20], Ontology-based Category Utility [56]**Error! Reference source not found.**, etc.

Variety of solutions [10]**Error! Reference source not found.** ranging from manual assignment by librarians [57]**Error! Reference source not found.** to unsupervised learning are provided for AND. Most of the researchers categorize $AND_{BD}$ in supervised, un-supervised and semi-supervised learning methods. The graph-based and ontology-based methods have also been applied to resolve AND. We have classified methods for AND in following five categories. Each category is explained in chorological order with discussions about their pros and cons.

### 6.1. Supervised Learning Methods

In supervised learning [19][24][28][30][58]-[60], the major objective is to find class labels by exploiting the related information. Supervised learning is labor intensive, costly and error-prone if labeling or training of the dataset is not performed properly. Supervised learning methods achieve better performance as compared to those of un-supervised learning methods with the tradeoff of expensive labelling labor and time consumed. Supervised methods may be exploited to predict an author name in a citation [24] or to disambiguate publications of a particular author [19][28][58][59].

Han et al. [24] proposed two supervised methods to disambiguate author names in the publications using VSM [45][61] for representation of publications; and cosine similarity for calculating pair-wise similarity of publication attributes. They propose canonical names by grouping together author names with the same first name initial and the same last name. Each canonical name is associated with all those publications, where that name appeared. First method applies naive Bayes probability model [62] and the second Support Vector Machines (SVMs) [63]. Both methods exploit triplet[1] attributes for similarity calculations. This famous work is actually the enhancement of Han et al. [64] where they exploited k-means clustering along with naïve Bayes model using the same dataset and attribute set.

Torvik et al. [28] proposed authority control framework to resolve only the name sharing problem for MEDLINE records by using eight different attributes. They calculated the pair-wise similarity profile on the basis of these attributes and decide whether a pair of publications containing the same name of an author belongs to the single individual. Culotta et al. [58] proposed a method that overcomes the problem of transitivity produced due to pair-wise comparisons. A researcher can have multiple papers, email addresses and affiliations. While comparing the publications of such authors the pair-wise classifier cannot handle multiple instances of an attribute. They employed the sets rather than pair-wise comparisons, and addressed the transitivity issue between co-authors in a better way. The comparison of a new publication is made with all the publications in a cluster rather than the pair-wise comparisons. By comparing a publication with sets makes it possible to handle the multiple values of an attribute.

Yin et al. [13] focused name sharing problem by considering only identical names. DISTINCT, an object distinction methodology to disambiguate authors is proposed. They combine set resemblance of neighbor tuples and random walk probability between the two records of relational database. SVM [63] is applied to assign weights to various types of links in the graph and agglomerative hierarchical clustering to get final clusters.

Torvik and Smalheiser [59] enhance their work [28] by (a) including first name and its variants, emails, and correlations between last names and affiliation words; (b) employing new procedures of constructing huge training sets; (c) exploiting methods for calculating the prior probability; (d) correcting transitivity violations by a weighted least squares algorithm; and (e) using agglomerative algorithm based on maximum likelihood for calculating clusters of articles that represent authors. The work proposed in [28] was not scalable which is usually a problem of most AND methods. The above enhancements make it scalable for a huge dataset like MEDLINE records.

---

[1] In this article we refer co-authors, title and venue attributes as triplet attributes.

-

Pucktada and Giles [19] resolve the name sharing problem in MEDLINE records. They introduce Random Forest classifier to find high-quality pair-wise linkage function. They define similarity profile by considering 21 attributes categorizing them in six types of attributes; three of them are triplets and other three are: affiliation similarity, concept similarity and author similarity. They use a naive based blocking procedure. This procedure uses author's last name and the first initial to block author name that does not share both parts of the author's name. They compare the results with SVM. Their results show that Random Forests outperforms SVM.

Qian et al. [60] proposed Labeling Oriented Author Disambiguation (LOAD) to resolve author name disambiguation problem together with users. LOAD exploits supervised training for estimating similarity between publications using High Precision Clusters (HPCs) for each author to change the labeling granularity from individual publications to clusters. Labeling HPCs decreases labeling effort at least 10 times as compared to the labeling publications. Found HPCs are clustered into High Recall Clusters (HRCs) to place all publications of one author into the same cluster. For pair-wise comparisons LOAD employs rich features like name, email, affiliation, homepage between two authors, co-author name, co-author email, co-author affiliation, co-author homepage, title bigram, reference and download link. Besides, self-citation and publishing year, interval between two papers are also considered.

The methods discussed above perform name disambiguation in offline environment. Different from them, Sun et al. [65] proposed publication analysis system. The focus of the system was to decide, at query time by involving user, if the queried author name matches the given set of publications retrieved from Google Scholar database. The system exploits two kinds of heuristic features (1) number of publications per name variation, and (2) publication topic consistency. Topic consistency exploits discipline tags crowd-sourced from the users of the Scholarometer system [66]. They train the binary classifier on a dataset of 500 top ranked authors from scholarometer database[1] by manually labeling either ambiguous or unambiguous, and examine the publications retrieved from Google Scholar for each queried name. To the best of our knowledge this is the first work addressing real-time author name disambiguation, and achieves 75% accuracy.

Zhang et al. [67] proposed a Bayesian non-exhaustive classification method for resolving online name disambiguation problems. They considered a case study for bibliographic data and involved a temporal stream format for disambiguating authors by dividing their papers into similar groups. Table 5 provides a quick summary of the methods based on supervised learning models.

Table 5: Summary of Supervised learning methods

| **Reference #** | **Problem** | **Tool / Method** | **Attributes / features** | **Comparison with** | **Dataset** | **Finding** | **Limitation** |
|---|---|---|---|---|---|---|---|
| **Han et al.** [24] **2004** | Disambiguate names in citations | naive Bayes probability model, SVM | co-author names, paper title, venue | Comparison of both approaches and their hybrid approach | Publications from web, DBLP | Hybrid of naive Bayes outperforms Hybrid I scheme of SVM | Not flexible, not topic sensitive |
| **Torvik et al.** [28] **2005** | resolve name sharing | authority control framework | 8 different attributes | Comparison is performed with manually labelled data only | Medline | Different articles authored by the same individual will share similarity in one or more aspect of Medline records | No comparison with state-of-the-art, Specific to Medline records only |
| **Culotta et al.** [58] **2007** | transitivity due to pair-wise comparisons | supervised machine learning, error-driven, rank-based training | Examining sets of records not pairs | Approach is evaluated on three different datasets | Penn, Rexa, DBLP | Error reduction of 60% over standard binary classification approach | Not topic sensitive, Not compared with state-of-the-art |
| **Yin et al.** [13] **2007** | name sharing problem | Supervised and un-supervised set resemblance and random walk | Fusion of different type of subtle linkages | Comparison of both approaches and their hybrid approach | DBLP | Fusing difference type of linkages and combining set resemblance of neighbor tuples and random walk probability is effective | Not compared with state-of-the-art, Specific to authors with identical name only |

[1] scholarometer.indiana.edu

| | | | | | | | |
|---|---|---|---|---|---|---|---|
| **Torvik and Smalheiser** [59] **2009** | Enhancement of [23] | estimating the probability that two articles sharing same name, were written by same individual | Adding 5 more variants to [23] | [23] | Medline | Author-ity model with more scalability and recall | Not high performance, model will fail to apply to scientists whose research output is diverse |
| **Pucktada and Giles** [19] **2009** | Name sharing problem | Random Forest classifier, naive based blocking | 21 different attributes | SVM | Medline | Random Forest classifier outperforms SVM | High accuracy can be achieved with a relatively small set of features. |
| **Qian et al.** [60] **2011** | Labeling Oriented Author Disambiguation | estimating similarity between publications using High Precision Clusters | Set of rich features | human labeling after conventional automatic author disambiguation | CS, UE and DBLP | Machine Learning combined with ceiv judgement produce more accurate results to assist and reduce human labeling | No Iterative process for AND, Limited usage of feature sources, Non usage of direct optimization algorithms |
| **Sun et al.** [65] **2011** | detect ambiguous names at query time | Finding ambiguities from crowdsourced annotations | number of citations per name variation, publication topic consistency | For each combination of features, accuracy, area under curve and F1 | Papers retrieved from google scholar | Improved accuracy | Publication metadata was not considered |
| **Zhang et al.** [67] **2016** | online name entity disambiguation | Dirichlet process prior with a Normal × Normal × Inverse Wishart data model | temporal stream format | Qian's Method [63], Khabsa's method [64] | Arnetminer | Proposed method outperforms the state-of-the-art methods | Computational complexity depends upon a number of factors and can be variable |

## 6.2. Unsupervised Learning Methods

Unsupervised learning methods [6][11][12][16][32][33][43][68]-[73] do not need manual labeling. Instead they carefully choose features to classify similar entities into clusters. Various clustering algorithms are applied to cluster the similar entities. Glies et al. [11] apply a k-way spectral clustering method to resolve AND. Unsupervised learning methods save labeling time with the tradeoff of efficiency and precision. However, in many dynamic scenarios unsupervised learning methods are better solution than the supervised learning methods.

The unsupervised methods may utilize similarities between publications with the help of predefined set of similarity functions to group the publications for a particular author. These functions are usually defined over the features present in the publications [11][12][32][68]-[71]. These features are also called the local information [17] as they are apparently available in the publication. The similarity functions may also be defined over implicit information such as topics of the publication [13][17][33] or Web data [33][72][73]. The information about the topic(s) of the publication is not explicitly present in the publication under consideration rather it is derived from the dataset hence called the global information [17].

Glies et al. [11] improved their previous work [24] by applying k-way spectral clustering [11] for AND using the triplet attributes for similarity measuring. Malin [12] applied hierarchical clustering and random walk to resolve name sharing and name variant problems. Main limitation of this method is a static threshold which is used as a stopping criteria of the clustering process. Bekkerman and McCallum [43] resolve name ambiguity problem. They present two frameworks: first one uses link structure of the Web pages, and second exploits A/CDC (Agglomerative / Conglomerative Double Clustering). Their methods require minimum of the prior knowledge as provided in BD. However their methods best fit to web appearances instead of BD.

Bhattacharya and Getoor [16] referred AND as entity resolution problem and extend LDA topic model [29]. They suppose that authors who belong to one or more groups of authors, may co-author papers and simultaneously discovered the clusters of authors and clusters of papers written by these authors. They perform parameter estimation through Expectation Maximization (EM) algorithm along with Gibbs sampling [74]. The extended model is about 100 times slower than an alternative method [69], and solves only the name variant problem. Bhattacharya and Getoor [69] proposed collective entity resolution method an improvement to their previous work [16]. Given two papers both written by authors $a_1$ and $a_2$, if the two instances of $a_2$ refer to the same individual, then it is likely that both instances of $a_1$ refer to the same entity. Resolving this 2$^{nd}$ level ambiguity helps in cases where there is a high level of ambiguity. They treat high verses low ambiguity scenarios separately. They first address the most confident assignments and then less confident ones. The final similarity value between the two citations is calculated on the basis of pair-wise comparisons and previously disambiguated authors. The weighting parameter is adjusted manually and it may take different optimal values across different contexts. Although this method is advancement to their previous work [16] yet scalability was still a problem.

Cota et al. [70] proposed a heuristic-based hierarchical clustering which successively combines clusters of citation records of the ambiguous authors. In first step, the compatibility of the ambiguous author names was found. If the two names in two publications are compatible then they are further compared against common compatible co-author(s). The two publications are merged to a cluster if compatible co-author is found, else they form separate clusters. The resulting clusters are almost pure but fragmented. To decrease the fragmentation they use second step in which clusters are compared in pair-wise fashion exploiting title and venue attributes. The major distinction of this method was that it compares all the titles and venues of a cluster with that of other clusters applying bag of words approach. If the similarity between two clusters reaches a threshold value then they are fused to one cluster otherwise they remain separate clusters. They claim improvements up to 12% against non-hierarchical clustering, 21% against SVM and 15.5% against K-means using the same attributes.

Song et al. [6] proposed an algorithm based on Probabilistic Latent Semantic Analysis [46] and Latent Dirichlet Allocation [29] to deal with AND exploiting contents of the articles. They exploited metadata of publications and authors and publication's first page to relate authors to topics.

Shin et al. [75] proposed AND framework by constructing social network for finding semantic relationships between authors and solves name sharing and name variant problems simultaneously. They employ two methods; one for namesake names and the other for heteronymous names. Social network is constructed in three steps. (1) *Information extraction:* extraction of paper title, etc. (2) *Candidate topics extraction:* extraction of topics that are representative of the publication. These candidate topics are extracted from abstract of the publication using morphemic analysis [76]. (3) *Social network construction*: the social network is constructed on the basis of above two types of information. They used the cosine similarity metric for finding similarity among two social networks.

Yang and Wu [77] resolves name sharing problem by exploiting triplet attributes along with web attribute. They use Cosine and Modified Sigmoid Function (MSF) for triplet attributes, and Maximum Normalized Document Frequency (MNDF) for web attribute, to estimate the pair-wise similarity between the publications. They also employed a binary classifier to reduce the noise in the clustering publications.

Tang et al. [7] formalize the problems for name disambiguation in a unified probabilistic framework. The framework uses a Markov Random Fields (MRF) [78] exploiting six local (publication) attributes (content based information) and five relationships (structure based information) between the pair of publications. The framework, on one hand, achieves better accuracy than baselines but, on the other hand, its time complexity is almost twice as compared to baselines.

Wu et al. [79] used Dempster-Shafer theory (DST) for AND. They proposed an unsupervised DST based hierarchical agglomerative clustering algorithm which is used with a combination of Shannon's entropy to blend disambiguation attributes for more reliable candidate pair of clusters for union in each repetition of clustering. Qian et al. [80] proposed a dynamic method for author name disambiguation keeping the growing nature of digital libraries in mind. They proposed a two-step process, BatchAD+IncAD, which first performs AND by grouping all records into disjoint clusters, and then it periodically performs incremental AND for newly added papers and determines that new papers belongs to an existing cluster or forms a new one. Khabsa et al. [81] proposed a constraint-based clustering algorithm, that allows constraints to be added to the clustering process and allowing the data to be added as well, in an incremental way. This methodology helps the users by allowing them to make corrections to disambiguated results. The method is based on a combination of DBSCAN and pairwise distance based on random forests. Sun et al. [82] proposed an unsupervised method based on topological features AND solution. To measure the similarity of publications the method includes a structure similarity algorithm along with random walk with restarts. Table 6 includes a summary of methods that involve unsupervised learning methods for AND.

Table 6: Summary of Unsupervised learning methods

| Reference # | Problem | Tool / Method | Attributes / features | Comparison with | Dataset | Finding | Limitation |
|---|---|---|---|---|---|---|---|
| **Glies et al.** [11] **2005** | Disambiguation in Author Citations | K-way Spectral Clustering | co-author names, paper titles, and publication venue titles | Evaluation based on confusion matrix | DBLP | spectral methods outperform k-means | Not compared with any state-of-the-art |
| **Malin** [12] **2005** | name sharing and name variant problems | hierarchical clustering and random walk | actor lists for movies and television shows | Consideration as baseline 1)ambiguous names are distinct entities 2) ambiguous names are single entity | IMDB | measuring similarity based on community, rather than exact similarity is more robust | Not compared with any state-of-the-art |
| **Bekkerman and McCallum** [43] **2005** | Finding Web appearances of a group of people. | link structure of the Web pages, another using Agglomerative/Conglomerative Double Clustering (A/CDC) | Only affiliation of a person with a group is required | traditional agglomerative clustering | hand-labeled a dataset of over 1000Web pages | Improved F measure | relational structure of relevant classes is not considered |
| **Bhattacharya and Getoor** [16] **2006** | Entity resolution | probabilistic model, extended LDA | decisions not on independent pairwise basis, but made collectively | hybrid SoftTF-IDF [31] | CiteSeer, arXiv (HEP) | exploits collaborative group structure for making resolution decisions | Cannot resolve multiple entity classes |
| **Bhattacharya and Getoor** [69] **2007** | Entity resolution | relational clustering algorithm | attribute-based baselines | attribute-based entity resolution, naïve relational entity resolution, collective relational entity resolution | CiteSeer, arXiv, BioBase | Improved performance over baselines | Manually adjusted weighting parameter which can have different optimal values. Not scalable |
| **Cota et al.** [70] **2007** | Disambiguation in split citation and mixed citation | heuristic-based hierarchical clustering | authors, title of the work, publication venue | SVM, K-Means | DBLP | Improved performance over baselines | Compared only with unsupervised methods |
| **Song et al.** [6] **2007** | disambiguation exploiting contents of the articles | Two stage approach based on LDA and PLSA | person names within web pages and scientific documents | spectral clustering and DBSCAN | CiteSeer | Improved scalability | Compared only with unsupervised methods |
| **Shin et al.** [75] **2010** | finding semantic relationships between authors and name sharing | Methods for namesake names and heteronymous names | Paper titles and topics | Comparison among two social networks with cosine similarity | DBLP | Improved effectiveness | -- |
| **Yang and Wu** [77] **2011** | name sharing problem | Cosine, Modified Sigmoid Function, and Maximum Normalized Document Frequency | triplet attributes along with web attribute | Compared with [11] | DBLP Dataset constructed by [11] | Improved accuracy | cluster separator filtered out some correctly matched pairs from the datasets |
| **Tang et al.** [7] **2012** | Disambiguation, how to find number of people "K" | Probabilistic Framework | Attributes of publications and relationships | Four baseline methods | ArnetMiner | Performs better than baseline and "K" is close to real | -- |
| **Wu et al.** [79] **2014** | Name disambiguation | DST based unsupervised hierarchical agglomerative clustring | | Three unsupervised models | | Performance comparable to a supervised model | -- |
| **Qian et al.** [80] **2015** | Dynamic disambiguation | BatchAD+IncAD framework | Authors metadata | five state-of-the-art batch AD methods | two labeled data sets, CaseStudy and DBLP | Improved efficiency and accuracy | Erroneous results when an author changes affiliation or topic |
| **Khabsa et al.** [81] **2015** | Disambiguation with constraints | DBSCAN and pairwise distance based on random forests. | metadata information and citation similarity | Models with different combination of features | CiteSeer | Improved pairwise and cluster F1 | DBSCAN cannot split an impure cluster |

### 6.3. Semi-Supervised Methods

Semi-supervised Learning approaches [31] have also been applied to AND in BD. It combines the characteristics of both supervised and unsupervised methods.

On et al. [26] proposed the frame work for resolving name variant problem in two steps: (1) blocking and (2) distance measurement. They used four blocking methods that reduce the candidates, and seven unsupervised distance measurements that measure the distance between the two candidate publications to decide whether they belong to the same entity. They also exploit two supervised algorithms Naive Bayes model [62] and the Support Vector Machines (SVMs) [63] to separate the publications of an author in a separate cluster.

Lee et al. [14] called the name sharing problem as mixed citation and name variant as split citation problem. They used Naive Bayes model and SVM (supervised methods); and cosine, TFIDF, Jaccard, Jaro and JaroWinkler (unsupervised methods) to resolve the name disambiguation problem.

On et el. [44] again focused on the name variant problem and call it Grouped-Entity Resolution (GER) problem. They propose Quasi-Clique, a graph partition based method. Unlike previous text similarity approaches like string distance, TFIDF or vector-based cosine metric, etc, their approach investigates the hidden relationship under the grouped-entities using Quasi-Clique technique.

Huang et al. [83] resolve both types of problems on a small dataset selected from CiteSeer. They employed an online SVM algorithm (LASVM) as supervised leaner of finding the distance metric of the publication attributes by pair-wise comparisons. The supervised learner easily handles the new papers with on line learning. For clustering the publications of the authors they used DBSCAN algorithm that constructs the clusters on multiple pair wise similarities and also handles the transitivity problem. They use different similarity metrics for different attributes, e.g., edit distance for URLs and emails, Jaccard similarity for affiliations and addresses, and Soft-TFIDF [84] for author names.

Zhang et al. [27] proposed a semi-supervised name disambiguation probabilistic model with six constraints. They consider following constraints: (1-3) triplet attributes constraints; (4) CoOrg, principal authors of two papers are from the same organization; (5) citation, one publication cites the other; (6) $\tau$-CoAuthor, two of the co-authors (one from each publication) are not same but they appear in another publication as co-authors. They applied Hidden Markov Random Fields for AND on arnetminer[1] data. Their model combines six types of constraints with Euclidean distance, and facilitates the user to refine the results.

Wang et al. [85] proposed a two-step semi-supervised method for AND that resolves name sharing problem only for identical names in Arnetminer[2]. They propose atomic clusters, i.e., each cluster has the publications of a particular author. At first step, they use a bias classifier to find the atomic clusters. They use a list of publications having the ambiguous author name and triplet attributes of the publications as input to the classifier. At second step, they integrate the atomic clustering results into the Hierarchical and K-means clustering algorithms.

Wang et al. [25] proposed constraint based topic modeling (CbTM) method as extension of [27]. They assume that if a pair of publications satisfies a constraint then both the publications should have more chances to have similar topic distribution. They combine the original likelihood function of LDA with a set of constraints defined over the attributes available from the publications' dataset. Thus the likelihood function is also affected by the constraints. They define the constraints as set of constraint functions each having value either 0 or 1. The presence of a constraint in the pair of publications under consideration means the function has value 1 otherwise 0. They define five constraints; two of them belong to triplet attributes excluding the title attribute and other three are: indirect co-author or transitive co-author (it is actually the $\tau$-CoAuthor constraint defined in [27]); web constraint (it means that two publications appear in the same web page) and user feedback (what the users comment about two publications' authors). At the end agglomerative hierarchical clustering algorithm is employed to construct clusters to uniquely identify authors containing all their publications.

Shu et al. [17] proposed LDA-dual topic model for complete entity resolution. They categorize AND into three types: name sharing, name variant and name mixing. They introduce the concept of global information based on the words and author names present in the dataset. In LDA-dual they define topics as two Dirichlet distributions, one over words and the other over author names, characterizing topics as series of words and author names. They also consider the local information like paper title and co-authors etc. Along with triplet attributes they use topic similarity and minimum name distance. They claim that two publications share little local information as compared to that of global information and employed Metropolis-Hasting within Gibbs sampling to calculate the global information i.e., model hyper parameters: α, β and γ. Complete process consisted of following steps: (1) find topics of publication in the dataset using Gibbs sampling; (2) construct a pair-wise classifier of two publications; (3) resolve name sharing problem with the help of spectral clustering and classifier's support for each ambiguous author name; (4) solve the name variant and name mixing problem with help of the classifier.

[1] http://arnetminer.org

Ferreira et al. [31] proposed Self-training Associative Name Disambiguation, a hybrid name disambiguation method. In the first (unsupervised) step clusters of authorship records are formed utilizing persistent patterns in the co-authorship graph. In the second (supervised) step training is performed through a subset of clusters constructed in the first step deriving the disambiguation function.

Arif et al. [86] proposed an enhanced version of vector space model for AND in digital libraries. Along with the normal authorship attributes, they added the additional information form the paper's metadata, including email ID, affiliation of authors and co-authors as well. These additional features have greatly improved the performance of the method. Table 7 shows the summary of name disambiguation methods that involve semi-supervised learning.

Table 7: Summary of Semi-supervised learning methods

| **Reference #** | **Problem** | **Tool / Method** | **Attributes / features** | **Comparison with** | **Dataset** | **Finding** | **Limitation** |
|---|---|---|---|---|---|---|---|
| **On et al.** [26] **2005** | name variant problem | (1) blocking and (2) distance measurement, 7 supervised and 2 unsupervised algorithms | Co-author relationships | four alternatives using three representative metrics | DBLP, e-Print, BioMed, EconPapers | using coauthor relation (instead of author name alone) shows improved scalability and accuracy | It is a two-step approach and shows improvement over one-step approach |
| **Lee et al.** [14] **2005** | Mixed citations and split citations | sampling-based approximate join algorithm, 2 supervised and 5 unsupervised | Associated information of author names | four alternatives using three representative metrics | DBLP, e-Print, BioMed, EconPapers | Improved accuracy | Accuracy for e-print is lower as compared to DBLP's accuracy |
| **On et el.** [44] **2006** | Name variant | graph partition based method Quasi-Clique | Contextual information mined from the group of elements | Quasi-Clique experimented on different real and synthetic datasets | ACM, BioMed, IMDB | improves precision and recall with existing ER solutions | Performance is better for IMDB but not for Citations data which has more strong connections as compared to actors in IMDB |
| **Huang et al.** [83] **2006** | name sharing, and name variant problem | LASVM and DBSCAN | Author and papers metadata | Traditional SVMs | CiteSeer | Improved efficiency and effectiveness | -- |
| **Zhang et al.** [27] **2007** | name disambiguation | semi-supervised probabilistic model | 6 different features from authors and citation information | Blocking and distance measure for co-authors | Arnetminer | improved scalability and accuracy | Compared only with unsupervised hierarchical clustering methods |
| **Wang et al.** [85] **2008** | name sharing problem | two-step semi-supervised method | Atomic clusters with citations of a particular author | Hierarchical clustering and K-means | Arnetminer | Concept of atomic clusters produce better results. Co-author features are important for atomic clusters | Compared only with unsupervised hierarchical clustering methods |
| **Shu et al.** [17] **2009** | name sharing, name variant and name mixing | LDA-dual topic model | generative latent topic model that involves both author names and words | experiments on three different training data sets | DBLP | Improved accuracy | smoothing method for new words and author names does not scale |
| **Ferreira et al.** [31] **2010** | Name disambiguation | Self-training Associative Name Disambiguation (SAND) | Authorship records | Two supervised and two unsupervised methods | DBLP, BDBComp | Improved results as compared to baselines | More improvement when compared with unsupervised methods as compared to the case of supervised methods |
| **Wang et al.** [25] **2010** | name sharing problem | constraint based topic modeling | combine the original likelihood function of LDA with a set of constraints | Hierarchical clustering algorithm to group the papers into clusters | Arnetminer | Improved precision, recall and F1 | -- |
| **Arif et al.** [86] **2014** | mixed citation and split citations problem | enhanced vector space model | additional attributes like e-mail ID and affiliation of author and co-authors | Comparisons of real authors names with names generated by proposed method | IEEE | Improved F measure | Not tested against any baseline or state-of-the-art |

### 6.4. Graph based Methods

The graph based methods are popular for AND. Many authors employ co-authorship graph to capture the similarity between two entities. It has been adopted by many methods discussed above, such as relational similarity in Bhattacharya and Getoor [69] and Yin et al. [13]; inter-object connection strength in Kalashnikov and Mehrotra [87], Yin et al. [13] and Chen et al. [88]; and semantic association in Jin et al. [89]. The length of the shortest path in a graph is usually employed to estimate the degree of closeness between two nodes. Kalashnikov and Mehrotra [87] and Yin et al. [13] utilized connection strength to find similarity of two nodes connected to each other through relationships. For this purpose Kalashnikov and Mehrotra [87] exploit legal paths and Fan et al. [20] make use of valid paths. Bhattacharya and Getoor [69] employed collaboration paths of length three and assign equal weights to all paths regardless of their length. Kalashnikov and Mehrotra [87] proposed more complicated method to calculate the weights for connection strengths. They proposed multiple equations and an iterative method to determine the weights. Differently, On et al. [44] used Quasi-Clique, a graph mining technique [90] to take advantage of contextual similarity in addition to syntactic similarity. On et al. [44], Chen et al. [88] and Jin et al. [89] estimate the similarity between two nodes (authors) as a combination of the feature-based similarity and the connection strength of the graph. Chen et al. [88] estimate the connection strength between two nodes as the sum of connection strengths of all simple paths no longer than a user-defined length.

In above paragraph we presented short but comparative description of some of the graph based works in AND. Now details of each work is discussed. McRae-Spencer and Shadbolt [91] resolved the AND on large scale citation networks through graph based method exploiting self-citation, co-authorship and publication source analyses in three passes to tie the papers of a particular author in a collection assigned to that author. First pass is to test each paper in the ambiguous name cluster against every other paper within that cluster to see if the second paper is the self-citation of the first, or vice versa. Similarly, the second pass is performed to draw a co-authorship graph, and the third pass used source URL metadata. The output of these three passes is the graphical representation of the publications. This method was based on metadata rather than textual context and on the notion that authors cite their own previous publications. As this method used self-citation as an attribute so the new papers have fewer or may have no citations at all. The papers of an author written just before his/her retirement[1] or death will never have self-citations. Similarly the papers written just before the change of research area will be self-cited hardly ever.

Galvez and Aneg´on [18] addressed the conflation of personal name variants problem in a standard or canonical form exploiting finite-state transducers and binary matrices. They divide the variants into *valid* (the variation among legitimate variants and canonical forms, e.g., such as the lack of some components of a full name, the absence or use of punctuation marks, the use of initials, etc) and *non-valid* (the variation among non-legitimate variants and correct forms, e.g., miss-spellings, involving deletions or insertions of characters in the strings, nicknames, abbreviations, errors of accentuation in the names from certain languages, etc) categories. They identify and conflate only valid variants into equivalence classes and canonical forms.

[1] By the term "retirement" we do not mean the retirement from job rather we mean retirement from research work willingly or unwillingly due to any reason.

Yin et al. [13] proposed DISTINCT, an object distinction methodology to solve AND, where entities have identical names. The method combines set resemblance of neighbor tuples and random walk probability (between two records in the graph of relational data) to measure relational similarity between the records of relational database. These two methods are complementary: one exploits the neighborhood information of the two records, and the other uses connection strength of linkages by assigning weights. DISTINCT exploits several types of linkages, like title, venue, publisher, year, authors' affiliation, etc.

Jin et al. [89] proposed Semantic Association AND graphical method. The similarity between the attributes (expect co-authors) of the two publications is measured through VSM, and the term TF-IDF is applied for term weighting. For co-authors and transitive co-authors semantic association graphs are constructed. The nodes show authors and the edges show the association. The edges also determine the weight by counting the number of publications co-authored by two authors. It is two-step process, RSAC (Related Semantic Association based Clustering) and SAM (Semantic Association based Merging). RSAC clusters two publications in a group if the co-authorship graphs of the two publications are similar, i.e., they have common co-authors. Similarly all the publications are grouped in small clusters. It is quite possible that transitivity property holds true for co-authors of some publications but RSAC does not handle it, and all the publications of an author may be assigned to multiple groups. To handle this issue SAM merges the groups on the basis of similarity values calculated for literatures (titles + abstracts), affiliations and transitive co-authorship graphs.

Fan et al. [20] resolved name sharing problem through GHOST (GrapHical framewOrk for name diSambiguaTion) using only co-authorship attribute, however for dense authors they exploited user feedback too. Contrary to the methods of Chen et al. [88] and Jin et al. [89], GHOST does not take into account the feature-based similarity, and the connection strength between nodes *u* and *v* is measured using *Ohm's Law*-like formula defined over subset of valid paths. Another difference of this work from the work in [89] is that it does not model the transitive co-authorship graph. This work has two strengths. First, the time complexity is very low as compared to the previous works as it exploits only co-author attribute and achieves 94% precision on average. Second, GHOST employs *Ohm's Law*-like formula to compute similarity between any pair of nodes in a co-authorship graph. The drawback of GHOST is that the results for dense authors are not in line with the results of non-dense authors. Fan et al. [20] proposed user feedback for such authors. No doubt the results are improved but the scalability is a challenge here because in real life databases there may be thousands of dense authors.

Wang et al. [61] proposed active user name disambiguation (ADANA) exploiting a pair-wise factor graph (PFG) model which can automatically determine the number of distinct names. Based on PFG model, they introduce a disambiguation algorithm that improves performance through user interaction.

Shin et al. [92] proposed a graph based model called Graph Framework for Author Disambiguation (GFAD), which involves co-author relations while constructing graphs and ambiguity is removed by vertex splitting and merging based on the co-authorship. Table 8 provides a summary of methods that involve use of graph based models.

Table 8: Summary of Graph based methods

| Reference # | Problem | Tool / Method | Attributes / features | Comparison with | Dataset | Finding | Limitation |
|---|---|---|---|---|---|---|---|
| **McRae-Spencer and Shadbolt** [91] **2006** | Name disambiguation | Citation graph | self-citation, co-authorship and document source analyses | Precision , recall an dF1 for 8 name based clusters | CiteSeer | Slightly improved results in terms of usefulness | Needs to create correction facility within some tested services |
| **Galvez and Aneg´on** [18] **2007** | personal name variants problem | standard or canonical form exploiting finite-state transducers and binary matrices | Author names | Application of master graph to the lists of author indexes | LISA, SCI-E. | Improved precision, Recall and F1, reduced erroneous analysis | similarity measures needs improvement in terms of error margins |
| **Jin et al.** [89] **2009** | Name disambiguation | semantic Association based Name Disambiguation method (SAND), | Semantic association graphs | DISTINCT [13], AKTiveAuthor [91] | CitesSeer, DBLP, Libra | Improved accuracy | -- |
| **Fan et al.** [20] **2011** | Name disambiguation | Graphical framework for name disambiguation (GHOST) | feature-based similarity, and the connection strength between nodes based on co-authorship | 2 labeled authors for DBLP and 8 labeled authors for PubMed for comparison, DISTINCT [13] | DBLP, PubMed | High precision and recall | performance may suffer for rare dense authors |
| **Wang et al.** [61] **2011** | Active name disambiguation | ADANA using pair-wise factor graph | active user interactions | 4 baseline methods | publication data set, a web page data set, and a news page data set | Reduced error rate | Error rate has been decreased with the help of user corrections |
| **Shin et al.** [92] **2014** | namesake problem | Graph Framework for Author Disambiguation | co-author relations | 3 representative unsupervised methods | DBLP, Arnetminer | Improved performance | -- |

### 6.5. Ontology based Methods

In information science, ontology is basically the knowledge of concepts and the relationships between those concepts within a domain. In other words, it is knowledge representation of a domain. Ontology based AND has been exploited by many researchers in different fields. For example, Geographic Named Entity Disambiguation [93], Identity Resolution Framework (IdRF) [94], Named Entity Disambiguation exploiting Wikipedia [95][96], Entity Co-reference [92], etc. As far as digital libraries or BD are concerned, researchers paid little attention to this kind of methods.

Initially, Hassell et al. [97] resolved AND through already populated ontology extracted from the DBLP. They utilize a file from DBLP that contains entities like authors, conferences and journals, and convert it into RDF and used it as background knowledge. Their method takes a set of documents from DBWorld[1] posts, "call for papers" to disambiguate the authors. Each such document contains multiple authors, say, the committee members, and some information about them, like affiliation; and information about the venue like topics of the venue. The scenario of the method is different from those we have discussed throughout this article. All other approaches perform disambiguation by either predicting the most probable author of a publication or by grouping the publications of the same author in a unique cluster in BD. Different from those, this method pinpoints, with high accuracy, the correct author in the DBLP ontology file that a document of DBWorld refers to. Their method selects an author name from the document and searches the candidate authors in the populated ontology in RDF form. All the candidate authors are compared with the author in the document to predict the most confident author in the ontology that relates to the author in the document. Different types of relationships in the ontology are exploited to predict the correct author out of various matches (candidates) in the ontology. These relationships include entity name, text proximity, text co-occurrence, popular entities and semantic relationships. Name entity refers to specifying which entities from the populated ontology are to be spotted in the text of the document and later disambiguated as all the entities of the document may not present in the DBLP ontology. Text proximity is the number of space characters between the name entity and the known affiliation. Here known affiliation means the object already known by the ontology as affiliation, say, name of a university. In DBWorld postings affiliations are usually written next to the entity name. If an entity name in the document and the affiliation matches the author name and known affiliation in the ontology, there are chances that these two entities are the same real world entity. Text co-occurrence is utilized to match the research areas of the candidate authors in the ontology and the topics of the venue present in the posting. Popular entity is the author in the ontology that has the highest score of publications among the candidate authors. Semantic relationships are used to match the co-authors of the candidate authors in the ontology and the entities in the document, with a notion that the entities on a document may be related to one another through any means, may be co-authors of some publications.

Park and Kim [56] proposed OnCu System to resolve name sharing problem through ontology-based category utility. The term category utility is used for similarity measurement between two entities. They exploit two types of ontology: *author ontology*, built on the publications from several proceedings of conferences, and the computer science *domain ontology*. Different from Hassell et al. [97] they determine the correct author from various candidate authors in the author ontology by exploiting the domain ontology for estimating the semantic similarity. Their goal is to discover the right author of the input publication and his/her right homepage. Their method also differs from that of Hassel et al. [97] in using ontology-based evaluation functions. OnCU views candidate authors as clusters of their publications and employs a cluster-based evaluation function exploiting ontology to predict the right author out of multiple candidate authors. The ontology-based approaches provided better semantic similarity measures for different attributes but this is fruitful only if the ontologies providing background knowledge are carefully constructed and frequently revised to meet the dynamic nature of the digital libraries. Table 9 provides a quick summary of disambiguation based that utilize the domain ontology.

Table 9: Summary of Ontology based methods

[1] DBWorld. http://www.cs.wisc.edu/dbworld/ April 9, 2006

| Reference # | Problem | Tool / Method | Attributes / features | Comparison with | Dataset | Finding | Limitation |
|---|---|---|---|---|---|---|---|
| **Hassell et al.** [97] **2006** | Entity disambiguation | Ontology-driven method | background knowledge (authors, conferences and journals) | Different types of relationships in the ontology are exploited | Ontology from DBLP, corpus from DBWorld | Successful use of large, populated ontology | Needs to be tested on more robust platforms |
| **Park and Kim** [56] **2008** | name sharing problem | OnCu, ontology-based category utility | author ontology, Computer science domain ontology | Evaluation based on category utility over the created ambiguity dataset | Collected papers from AAAI, ISWC, ESWC, And WWW conferences websites. | Improved performance | Cannot consider property relations |

## 7. Performance Evaluation

Accuracy, precision, recall and F-measure are the common performance metrics used to evaluate AND methods [7][16][17][20][25][27][43][61][75]. Performance of method used is either measured in terms of number of publications correctly predicted or the number of authors correctly predicted. In literature, the performance measurement terms are defined in variety of ways. Here we shortly describe the common notion of these terms:

### 7.1. Accuracy

Accuracy (disambiguation accuracy) is the generic term used to represent performance in terms of correctness. It may be defined in any way that best suits the proposed method. It may be equivalent to precision, recall, F-measure, etc. The term accuracy is defined and used by several researchers [14][19][24][30]. For example, Han et al. [24] defined disambiguation accuracy as "the percentage of the query names correctly predicted", whereas Han et al. [30] defined it as "the sum of diagonal elements divided by the total number of elements in the confusion matrix". Both these definitions describe the accuracy in terms of correctly predicted authors rather than the correctly predicted publications of an author.

### 7.2. Precision

It is the ratio between the number of correctly predicted publications of author $a_i$ and the number of publication predicted as $a_i$*'s* publications.

$$Precision = \frac{No.of\ elements\ of\ [\{P_{a_i}\} \cap \{P'_{a_i}\}]}{No.of\ elements\ of\ \{P'_{a_i}\}} -------- (7)$$

where, $P_{a_i}$ = publications of author $a_i$ and $P'_{a_i}$ = publications predicted as author $a_i$*'s*. Suppose, author $a_i$ has publications {$P_1$-$P_5$}; and the system predicted publications of author $a_i$ are {$P_1$-$P_4$, $P_6$, $P_7$}. By applying Eq. 7:

Precision = 4/6 = 0.67

### 7.3. Recall

It is the ratio between the number of correctly predicted publications of author $a_i$ and number of $a_i$*'s* publications.

$$Recall = \frac{No.of\ elements\ of\ [\{P_{a_i}\} \cap \{P'_{a_i}\}]}{No.of\ elements\ of\ \{P_{a_i}\}} -------- (8)$$

where, $P_{a_i}$ = Publications of author $a_i$ and $P'_{a_i}$ = Publications predicted as author $a_i$*'s*. By considering the above example using Eq. 8:

Recall = 4/5 = 0.8

### 7.4. F-Measure

It is the harmonic mean of precision and recall.

$$H = \frac{n}{\sum_{i=1}^{n} \frac{1}{x_i}} -------- (9)$$

By consider the above example using Eq. 9:

$$\text{F-measure} = \frac{2}{\left(\frac{1}{0.67} + \frac{1}{0.8}\right)} = \frac{2}{(1.49 + 1.25)} = 0.73$$

The above metrics can also be defined on the cluster level too [31]. Cluster precision is the fraction of correct clusters to the number of clusters acquired by the method, and cluster recall is the fraction of true clusters to that of the method, and cluster F-measure is the harmonic mean of the both [31].

## 8. Future Directions and Recommendations

Although a lot of research work has been performed in this field yet there is a need for a lot of improvement. Many attempts have been made to assign a unique author ID to each author to resolve the name disambiguation but these methods could not gain the attention of the researchers due to many reasons as we have discussed in Section 2. Many researchers emphasize to exploit more and more attributes to estimate the maximum similarity among the citations. This causes two issues: *first*, the time complexity of the algorithm increases and resultantly scalability is inversely affected; *second*, the availability of numerous features for each citation becomes almost impossible. Besides these issues assigning weight and fixing threshold values to each feature are bottleneck, especially when the feature set becomes large. We recommend exploiting only those features that are usually available in the BD so that a general framework applicable to most of them can be proposed. To resolve the AND problem in a better way we suggest few directions that may help improve the performance.

#### 8.000 Structured and Un-Structured Semantics

Semantics plays and important rule in co-author networks [98][99][100]. WordNet[1] captures structured semantics of words and can be exploited for AND in BD to achieve more accurate results through ontologies [56,97]. Un-Structured semantics [][]We propose to use multi-gram topic models beside the unigrams of words for topics' distribution over words. In this way, the natural syntactic relationship among the words is preserved and author writing habits can become useful for AND.

First and second suggestions are useful as they consider semantics and can provide better similarity estimation among the citations.

Third: In literature, the transitivity issue is addressed only for co-authors attribute. We suggest leveraging this concept for title and venue attributes too. Fourth: Instead of simply matching the titles of the publications the references of the two publications to find the similarity between the two publications can also be exploited. Fifth: most of the methods while handling the venue attribute use only its title. We suggest considering the ranking of the publication venues too. On the basis of this ranking, the REsearch Ability Level (REAL) of a researcher can be estimated. The REAL value may help predict the correct author as authors with same names might have different rank research publications. All these measurements (first to fifth) help improve the similarity metrics. Sixth: The change of research domain of an author is common these days due to overlaps between different fields. We suggest constructing sub-clusters within a cluster associated to a particular author. Each sub-cluster can be differed from those of others on the basis of multiple topics of interest of the author. Seventh: The advisor-advisee relationship can also be found first to develop hierarchies for authors resultantly the authors which are not same will become nodes in different stems of a tree.

## 9. Conclusions

In this survey, we presented a detailed study of the AND methods for DB. Key challenges are highlighted and generic framework is proposed, which is quite intuitive and applicable. A lot of work has been done for name variant and name sharing problem separately, but few efforts are made to deal both of them simultaneously which needs more attention. Different types of methods, such as, supervised, up-supervised, semi-supervised, graph based and ontology based provided elegant solutions for AND, still graph based and ontology based methods needs to be explored exhaustively. At the end we have highlighted the major issues and future directions in this field. These future directions and open challenges can give a quick start to future researchers who are interested to do research in this area.

As a whole in this paper, we presented a snapshot of research work done about AND in BD, methods applied and future challenges around the time of its writing. However, we do believe that the fundamental information, methods, future directions and open challenges presented here will be useful for the researchers in this area of research now and in future to get a quick start.

[1] http://wordnet.princeton.edu/

**Acknowledgement**

We are grateful to the Higher Education Commission (HEC) of Pakistan for their financial assistance to promote the research trend in the country under Indigenous 5000 Fellowship Program.

**References**

[1] T. Amjad, A. Daud, D. Che, and A. Akram, "MuICE: Mutual Influence and Citation Exclusivity Author Rank," *Inf. Process. Manag.*, 2015.

[2] T. Amjad, A. Daud, A. Akram, and F. Muhammed, "Impact of mutual influence while ranking authors in a co-authorship network," *Kuwait J. Sci.*, vol. 43, no. 3, 2016.

[3] C. Laorden, I. Santos, B. Sanz, G. Alvarez, and P. G. Bringas, "Word sense disambiguation for spam filtering," *Electron. Commer. Res. Appl.*, vol. 11, no. 3, pp. 290–298, 2012.

[4] J. Miró-Borrás and P. Bernabeu-Soler, "Text entry in the e-commerce age: two proposals for the severely handicapped," *J. Theor. Appl. Electron. Commer. Res.*, vol. 4, no. 1, pp. 101–112, 2009.

[5] D. Chen, X. Li, Y. Liang, and J. Zhang, "A semantic query approach to personalized e-Catalogs service system," *J. Theor. Appl. Electron. Commer. Res.*, vol. 5, no. 3, pp. 39–54, 2010.

[6] Y. Song, J. Huang, I. G. Councill, J. Li, and C. L. Giles, "Efficient topic-based unsupervised name disambiguation," in *Proceedings of the 7th ACM/IEEE-CS joint conference on Digital libraries*, 2007, pp. 342–351.

[7] J. Tang, A. C. Fong, B. Wang, and J. Zhang, "A unified probabilistic framework for name disambiguation in digital library," *IEEE Trans. Knowl. Data Eng.*, vol. 24, no. 6, pp. 975–987, 2012.

[8] M. Ley, "The DBLP computer science bibliography: Evolution, research issues, perspectives," in *International Symposium on String Processing and Information Retrieval*, 2002, pp. 1–10.

[9] C. L. Giles, K. D. Bollacker, and S. Lawrence, "CiteSeer: An automatic citation indexing system," in *Proceedings of the third ACM conference on Digital libraries*, 1998, pp. 89–98.

[10] N. R. Smalheiser and V. I. Torvik, "Author name disambiguation," *Annu. Rev. Inf. Sci. Technol.*, vol. 43, no. 1, pp. 1–43, 2009.

[11] C. L. Giles, H. Zha, and H. Han, "Name disambiguation in author citations using a k-way spectral clustering method," in *Proceedings of the 5th ACM/IEEE-CS Joint Conference on Digital Libraries (JCDL'05)*, 2005, pp. 334–343.

[12] B. Malin, "Unsupervised name disambiguation via social network similarity," in *Workshop on link analysis, counterterrorism, and security*, 2005, vol. 1401, pp. 93–102.

[13] X. Yin, J. Han, and S. Y. Philip, "Object distinction: Distinguishing objects with identical names," in *2007 IEEE 23rd International Conference on Data Engineering*, 2007, pp. 1242–1246.

[14] D. Lee, B.-W. On, J. Kang, and S. Park, "Effective and scalable solutions for mixed and split citation problems in digital libraries," in *Proceedings of the 2nd international workshop on Information quality in information systems*, 2005, pp. 69–76.

[15] Y. F. Tan, M. Y. Kan, and D. Lee, "Search engine driven author disambiguation," in *Proceedings of the 6th ACM/IEEE-CS joint conference on Digital libraries*, 2006, pp. 314–315.

[16] I. Bhattacharya and L. Getoor, "A Latent Dirichlet Model for Unsupervised Entity Resolution.," in *SDM*, 2006, vol. 5, p. 59.

[17] L. Shu, B. Long, and W. Meng, "A latent topic model for complete entity resolution," in *Data Engineering, 2009. ICDE'09. IEEE 25th International Conference on*, 2009, pp. 880–891.

[18] C. Galvez and F. Moya-Anegón, "Approximate personal name-matching through finite-state graphs," *J. Am. Soc. Inf. Sci. Technol.*, vol. 58, no. 13, pp. 1960–1976, 2007.

[19] P. Treeratpituk and C. L. Giles, "Disambiguating authors in academic publications using random forests," in *Proceedings of the 9th ACM/IEEE-CS joint conference on Digital libraries*, 2009, pp. 39–48.

[20] X. Fan, J. Wang, X. Pu, L. Zhou, and B. Lv, "On graph-based name disambiguation," *J. Data Inf. Qual. JDIQ*, vol. 2, no. 2, p. 10, 2011.

[21] L. K. Branting, "A comparative evaluation of name-matching algorithms," in *Proceedings of the 9th international conference on Artificial intelligence and law*, 2003, pp. 224–232.

[22] D. A. Dervos, N. Samaras, G. Evangelidis, J. Hyvärinen, and Y. Asmanidis, "The universal author identifier system (UAI_Sys)," 2006.

[23] A. M. Ketchum, "ORCID," 2014.

[24] H. Han, L. Giles, H. Zha, C. Li, and K. Tsioutsiouliklis, “Two supervised learning approaches for name disambiguation in author citations,” in *Digital Libraries, 2004. Proceedings of the 2004 joint ACM/IEEE conference on*, 2004, pp. 296–305.
[25] F. Wang, J. Tang, J. Li, and K. Wang, “A constraint-based topic modeling approach for name disambiguation,” *Front. Comput. Sci. China*, vol. 4, no. 1, pp. 100–111, 2010.
[26] B.-W. On, D. Lee, J. Kang, and P. Mitra, “Comparative study of name disambiguation problem using a scalable blocking-based framework,” in *Proceedings of the 5th ACM/IEEE-CS joint conference on Digital libraries*, 2005, pp. 344–353.
[27] D. Zhang, J. Tang, J. Li, and K. Wang, “A constraint-based probabilistic framework for name disambiguation,” in *Proceedings of the sixteenth ACM conference on Conference on information and knowledge management*, 2007, pp. 1019–1022.
[28] V. I. Torvik, M. Weeber, D. R. Swanson, and N. R. Smalheiser, “A probabilistic similarity metric for Medline records: A model for author name disambiguation,” *J. Am. Soc. Inf. Sci. Technol.*, vol. 56, no. 2, pp. 140–158, 2005.
[29] D. M. Blei, A. Y. Ng, and M. I. Jordan, “Latent dirichlet allocation,” *J. Mach. Learn. Res.*, vol. 3, pp. 993–1022, 2003.
[30] H. Han, W. Xu, H. Zha, and C. L. Giles, “A hierarchical naive Bayes mixture model for name disambiguation in author citations,” in *Proceedings of the 2005 ACM symposium on Applied computing*, 2005, pp. 1065–1069.
[31] A. A. Ferreira, A. Veloso, M. A. Gonçalves, and A. H. Laender, “Effective self-training author name disambiguation in scholarly digital libraries,” in *Proceedings of the 10th annual joint conference on Digital libraries*, 2010, pp. 39–48.
[32] B.-W. On and D. Lee, “Scalable Name Disambiguation using Multi-level Graph Partition.,” in *SDM*, 2007, pp. 575–580.
[33] K.-H. Yang, H.-T. Peng, J.-Y. Jiang, H.-M. Lee, and J.-M. Ho, “Author name disambiguation for citations using topic and web correlation,” in *International Conference on Theory and Practice of Digital Libraries*, 2008, pp. 185–196.
[34] P. Reuther, “Personal name matching: New test collections and a social network based approach,” *Comput. Sci. Tech. Rep.*, pp. 6–1, 2006.
[35] S. Pandit, S. Gupta, and others, “A comparative study on distance measuring approaches for clustering,” *Int. J. Res. Comput. Sci.*, vol. 2, no. 1, pp. 29–31, 2011.
[36] W. Cohen, P. Ravikumar, and S. Fienberg, “A comparison of string metrics for matching names and records,” in *Kdd workshop on data cleaning and object consolidation*, 2003, vol. 3, pp. 73–78.
[37] A. E. Monge, C. Elkan, and others, “The Field Matching Problem: Algorithms and Applications.,” in *KDD*, 1996, pp. 267–270.
[38] R. Durbin, S. R. Eddy, A. Krogh, and G. Mitchison, *Biological sequence analysis: probabilistic models of proteins and nucleic acids*. Cambridge university press, 1998.
[39] M. A. Jaro, “Probabilistic linkage of large public health data files,” *Stat. Med.*, vol. 14, no. 5–7, pp. 491–498, 1995.
[40] W. E. Winkler, “The state of record linkage and current research problems,” in *Statistical Research Division, US Census Bureau*, 1999.
[41] Y. Chen and J. Martin, “Towards Robust Unsupervised Personal Name Disambiguation.,” in *EMNLP-CoNLL*, 2007, pp. 190–198.
[42] L. Jin, C. Li, and S. Mehrotra, “Efficient record linkage in large data sets,” in *Database Systems for Advanced Applications, 2003.(DASFAA 2003). Proceedings. Eighth International Conference on*, 2003, pp. 137–146.
[43] R. Bekkerman and A. McCallum, “Disambiguating web appearances of people in a social network,” in *Proceedings of the 14th international conference on World Wide Web*, 2005, pp. 463–470.
[44] B.-W. On, E. Elmacioglu, D. Lee, J. Kang, and J. Pei, “Improving grouped-entity resolution using quasi-cliques,” in *Sixth International Conference on Data Mining (ICDM’06)*, 2006, pp. 1008–1015.
[45] G. Salton, A. Wong, and C.-S. Yang, “A vector space model for automatic indexing,” *Commun. ACM*, vol. 18, no. 11, pp. 613–620, 1975.
[46] T. Hofmann, “Probabilistic latent semantic analysis,” in *Proceedings of the Fifteenth conference on Uncertainty in artificial intelligence*, 1999, pp. 289–296.
[47] T. L. Griffiths and M. Steyvers, “Finding scientific topics,” *Proc. Natl. Acad. Sci.*, vol. 101, no. suppl 1, pp. 5228–5235, 2004.
[48] A. Daud, J. Li, L. Zhou, and F. Muhammad, “Knowledge discovery through directed probabilistic topic models: a survey,” *Front. Comput. Sci. China*, vol. 4, no. 2, pp. 280–301, 2010.

[49] M. A. Hernández and S. J. Stolfo, "The merge/purge problem for large databases," in *ACM Sigmod Record*, 1995, vol. 24, pp. 127–138.
[50] H. L. Dunn, "Record linkage*," *Am. J. Public Health Nations Health*, vol. 36, no. 12, pp. 1412–1416, 1946.
[51] D. Bitton and D. J. DeWitt, "Duplicate record elimination in large data files," *ACM Trans. Database Syst. TODS*, vol. 8, no. 2, pp. 255–265, 1983.
[52] K. J. Cios, R. W. Swiniarski, W. Pedrycz, and L. A. Kurgan, "The knowledge discovery process," in *Data Mining*, 2007, pp. 9–24.
[53] W. W. Cohen, H. Kautz, and D. McAllester, "Hardening soft information sources," in *Proceedings of the sixth ACM SIGKDD international conference on Knowledge discovery and data mining*, 2000, pp. 255–259.
[54] A. Bagga, *Coreference, cross-document coreference, and information extraction methodologies*. Duke University, 1998.
[55] H. Pasula, B. Marthi, B. Milch, S. Russell, and I. Shpitser, "Identity uncertainty and citation matching," in *Advances in neural information processing systems*, 2002, pp. 1401–1408.
[56] Y.-T. Park and J.-M. Kim, "OnCU system: ontology-based category utility approach for author name disambiguation," in *Proceedings of the 2nd international conference on Ubiquitous information management and communication*, 2008, pp. 63–68.
[57] C. L. Scoville, E. D. Johnson, and A. L. McConnell, "When A. Rose is not A. Rose: the vagaries of author searching," *Med. Ref. Serv. Q.*, vol. 22, no. 4, pp. 1–11, 2003.
[58] A. Culotta, P. Kanani, R. Hall, M. Wick, and A. McCallum, "Author disambiguation using error-driven machine learning with a ranking loss function," in *Sixth International Workshop on Information Integration on the Web (IIWeb-07), Vancouver, Canada*, 2007.
[59] V. I. Torvik and N. R. Smalheiser, "Author name disambiguation in MEDLINE," *ACM Trans. Knowl. Discov. Data TKDD*, vol. 3, no. 3, p. 11, 2009.
[60] Y. Qian, Y. Hu, J. Cui, Q. Zheng, and Z. Nie, "Combining machine learning and human judgment in author disambiguation," in *Proceedings of the 20th ACM international conference on Information and knowledge management*, 2011, pp. 1241–1246.
[61] X. Wang, J. Tang, H. Cheng, and S. Y. Philip, "Adana: Active name disambiguation," in *2011 IEEE 11th International Conference on Data Mining*, 2011, pp. 794–803.
[62] V. Vapnik, *The nature of statistical learning theory*. Springer Science & Business Media, 2013.
[63] N. Cristianini and J. Shawe-Taylor, *An introduction to support vector machines and other kernel-based learning methods*. Cambridge university press, 2000.
[64] H. Han, H. Zha, and C. L. Giles, "A model-based k-means algorithm for name disambiguation," in *Proceedings of the 2nd International Semantic Web Conference (ISWC-03) Workshop on Semantic Web Technologies for Searching and Retrieving Scientific Data*, 2003.
[65] X. Sun, J. Kaur, L. Possamai, and F. Menczer, "Detecting ambiguous author names in crowdsourced scholarly data," in *Privacy, Security, Risk and Trust (PASSAT) and 2011 IEEE Third Inernational Conference on Social Computing (SocialCom), 2011 IEEE Third International Conference on*, 2011, pp. 568–571.
[66] D. T. Hoang, J. Kaur, and F. Menczer, "Crowdsourcing scholarly data," 2010.
[67] B. Zhang, M. Dundar, and M. A. Hasan, "Bayesian Non-Exhaustive Classification A Case Study: Online Name Disambiguation using Temporal Record Streams," *ArXiv Prepr. ArXiv160705746*, 2016.
[68] E. Elmacioglu, J. Kang, D. Lee, J. Pei, and B. On, "An effective approach to entity resolution problem using quasi-clique and its application to digital libraries," in *Proceedings of the 6th ACM/IEEE-CS Joint Conference on Digital Libraries (JCDL'06)*, 2006, pp. 51–52.
[69] I. Bhattacharya and L. Getoor, "Collective entity resolution in relational data," *ACM Trans. Knowl. Discov. Data TKDD*, vol. 1, no. 1, p. 5, 2007.
[70] R. G. Cota, M. A. Gonçalves, and A. H. Laender, "A Heuristic-based Hierarchical Clustering Method for Author Name Disambiguation in Digital Libraries.," in *SBBD*, 2007, pp. 20–34.
[71] J. Soler, "Separating the articles of authors with the same name," *Scientometrics*, vol. 72, no. 2, pp. 281–290, 2007.
[72] I.-S. Kang *et al.*, "On co-authorship for author disambiguation," *Inf. Process. Manag.*, vol. 45, no. 1, pp. 84–97, 2009.
[73] D. A. Pereira, B. Ribeiro-Neto, N. Ziviani, A. H. Laender, M. A. Gonçalves, and A. A. Ferreira, "Using web information for author name disambiguation," in *Proceedings of the 9th ACM/IEEE-CS joint conference on Digital libraries*, 2009, pp. 49–58.
[74] A. E. Gelfand, "Gibbs sampling," *J. Am. Stat. Assoc.*, vol. 95, no. 452, pp. 1300–1304, 2000.

[75] D. Shin, T. Kim, H. Jung, and J. Choi, "Automatic method for author name disambiguation using social networks," in *2010 24th IEEE International Conference on Advanced Information Networking and Applications*, 2010, pp. 1263–1270.
[76] D. Shin, J. Kang, J. Choi, and J. Yang, "Detecting collaborative fields using social networks," in *Networked Computing and Advanced Information Management, 2008. NCM'08. Fourth International Conference on*, 2008, vol. 1, pp. 325–328.
[77] K.-H. Yang and Y.-H. Wu, "Author name disambiguation in citations," in *Proceedings of the 2011 IEEE/WIC/ACM International Conferences on Web Intelligence and Intelligent Agent Technology-Volume 03*, 2011, pp. 335–338.
[78] R. Kindermann and L. Snell, *Markov random fields and their applications*. 1980.
[79] H. Wu, B. Li, Y. Pei, and J. He, "Unsupervised author disambiguation using Dempster–Shafer theory," *Scientometrics*, vol. 101, no. 3, pp. 1955–1972, 2014.
[80] Y. Qian, Q. Zheng, T. Sakai, J. Ye, and J. Liu, "Dynamic author name disambiguation for growing digital libraries," *Inf. Retr. J.*, vol. 18, no. 5, pp. 379–412, 2015.
[81] M. Khabsa, P. Treeratpituk, and C. L. Giles, "Online person name disambiguation with constraints," in *Proceedings of the 15th ACM/IEEE-CS Joint Conference on Digital Libraries*, 2015, pp. 37–46.
[82] C.-C. Sun, D.-R. Shen, Y. Kou, T.-Z. Nie, and G. Yu, "Topological Features Based Entity Disambiguation," *J. Comput. Sci. Technol.*, vol. 31, no. 5, pp. 1053–1068, 2016.
[83] J. Huang, S. Ertekin, and C. L. Giles, "Efficient name disambiguation for large-scale databases," in *European Conference on Principles of Data Mining and Knowledge Discovery*, 2006, pp. 536–544.
[84] M. Bilenko, R. Mooney, W. Cohen, P. Ravikumar, and S. Fienberg, "Adaptive name matching in information integration," *IEEE Intell. Syst.*, vol. 18, no. 5, pp. 16–23, 2003.
[85] F. Wang, J. Li, J. Tang, J. Zhang, and K. Wang, "Name disambiguation using atomic clusters," in *Web-Age Information Management, 2008. WAIM'08. The Ninth International Conference on*, 2008, pp. 357–364.
[86] T. Arif, R. Ali, and M. Asger, "Author name disambiguation using vector space model and hybrid similarity measures," in *Contemporary Computing (IC3), 2014 Seventh International Conference on*, 2014, pp. 135–140.
[87] D. V. Kalashnikov and S. Mehrotra, "Domain-independent data cleaning via analysis of entity-relationship graph," *ACM Trans. Database Syst. TODS*, vol. 31, no. 2, pp. 716–767, 2006.
[88] Z. Chen, D. V. Kalashnikov, and S. Mehrotra, "Adaptive graphical approach to entity resolution," in *Proceedings of the 7th ACM/IEEE-CS joint conference on Digital libraries*, 2007, pp. 204–213.
[89] H. Jin, L. Huang, and P. Yuan, "Name disambiguation using semantic association clustering," in *e-Business Engineering, 2009. ICEBE'09. IEEE International Conference on*, 2009, pp. 42–48.
[90] J. Pei, D. Jiang, and A. Zhang, "On mining cross-graph quasi-cliques," in *Proceedings of the eleventh ACM SIGKDD international conference on Knowledge discovery in data mining*, 2005, pp. 228–238.
[91] D. M. McRae-Spencer and N. R. Shadbolt, "Also by the same author: AKTiveAuthor, a citation graph approach to name disambiguation," in *Proceedings of the 6th ACM/IEEE-CS joint conference on Digital libraries*, 2006, pp. 53–54.
[92] D. Shin, T. Kim, J. Choi, and J. Kim, "Author name disambiguation using a graph model with node splitting and merging based on bibliographic information," *Scientometrics*, vol. 100, no. 1, pp. 15–50, 2014.
[93] J. Kleb and R. Volz, "Ontology based entity disambiguation with natural language patterns," in *Digital Information Management, 2009. ICDIM 2009. Fourth International Conference on*, 2009, pp. 1–8.
[94] M. Yankova, H. Saggion, and H. Cunningham, "Adopting ontologies for multisource identity resolution," in *Proceedings of the first international workshop on Ontology-supported business intelligence*, 2008, p. 6.
[95] H. T. Nguyen and T. H. Cao, "Named entity disambiguation on an ontology enriched by Wikipedia," in *Research, Innovation and Vision for the Future, 2008. RIVF 2008. IEEE International Conference on*, 2008, pp. 247–254.
[96] H. T. Nguyen and T. H. Cao, "Enriching ontologies for named entity disambiguation," in *Proceedings of the 4th International Conference on Advances in Semantic Processing (SEMAPRO 2010)*, 2010.
[97] J. Hassell, B. Aleman-Meza, and I. B. Arpinar, "Ontology-driven automatic entity disambiguation in unstructured text," in *International Semantic Web Conference*, 2006, pp. 44–57.
[98] A. Daud, "Using Time Topic Modeling for Semantics-Based Dynamic Research Interest Finding," *Knowledge-Based Systems (KBS)*, vol. 26, pp. 154-163, 2012.
[99] A. Daud and F. Muhammad, "Group Topic Modeling for Academic Knowledge Discovery," *Journal of Applied Intelligence*, vol. 36, no. 4, pp. 876-886, 2012.
[100] A. Daud, J. Li, L. Zhou, and F. Muhammad, "Temporal Expert Finding through Generalized Time Topic Modeling," *Knowledge-Based Systems (KBS)*, vol. 23, no. 6, pp. 615-625, 2010.